\documentstyle[12pt]{article}
\newcommand{\beq}{\begin{equation}}
\newcommand{\eeq}{\end{equation}}
\newcommand{\bea}{\begin{eqnarray}}
\newcommand{\eea}{\end{eqnarray}}
\newcommand{\Veff}%
{{\cal V}^{\mbox{\scriptsize p}}_{\mbox{\scriptsize eff}}}
\newcommand{\inn}{{\mbox{\scriptsize in}}}
\newcommand{\exx}{{\mbox{\scriptsize ex}}}
\newcommand{\lsim}{\stackrel{\scriptstyle <}{\phantom{}_{\sim}}}
\newcommand{\gsim}{\stackrel{\scriptstyle >}{\phantom{}_{\sim}}}
\newcommand{\fermi}{{\mbox{fm}}}

\begin{document}

\centerline{\large\bf Surface behaviour of the pairing gap}
\centerline{\large\bf in semi-infinite nuclear matter}

\vskip 1 cm

\centerline{M.$\,$Baldo$^{1}$, U.$\,$Lombardo$^{2,3}$,
E.$\,$Saperstein$^{4}$ and M.$\,$Zverev$^{4}$}

\vskip 1 cm

\centerline{$^1$INFN, Sezione di Catania, 57 Corso Italia,
I-95129 Catania, Italy}
\centerline{$^2$INFN-LNS, 44 Via S.-Sofia, I-95123 Catania, Italy}
\centerline{$^3$ Dipartimento di Fisica, 57 Corso Italia,
I-95129 Catania, Italy}

\centerline{$^4$ Kurchatov Institute, 123182, Moscow, Russia }

\vskip 3 cm

\centerline
{\bf Abstract}
\vskip .3 cm

The $^1S_0$-pairing gap in semi-infinite nuclear matter is evaluated
microscopically using the effective pairing interaction
recently found explicitly in the coordinate representation
starting from the separable form of the Paris $NN$-potential.
Instead of direct iterative solution of the gap equation,
a new method proposed by
V.~A.~Khodel, V.~V.~Khodel and J.~W.~Clark
was used which simplifies the procedure
significantly. The gap $\Delta$ obtained in our calculations
exhibits a strong variation in the surface region
with a pronounced maximum near the surface.

\newpage

Recently  \cite{BLSZ1,BLSZ4} a method was elaborated of explicit
consideration of the gap equation for semi-infinite nuclear matter
within an approach based on the concept of the effective pairing
interaction.
The use of a separable form of the $NN$-potential simplifies the problem
significantly making it solvable numerically
in the mixed coordinate-momentum representation. In Ref.~\cite{BLSZ4}
the effective pairing interaction in the $^1S_0$-channel  was obtained
explicitly
without any form of local approximation for the separable $3\times 3$
form \cite{Par1,Par2} of the Paris potential \cite{PARIS}.
In this case the gap equation is reduced to a set of three integral
equations in coordinate space with kernels which are expressed
in terms of $u$- and $v$-functions obeying the Bogolyubov equations
 with nonlocal gap $\Delta$. These equations
have the integro-differential form that makes
their solution rather complicate.
In Ref.~\cite{BLSZ5} a method was elaborated of
solving such equations, but it proved out to be rather cumbersome.
As it is well-known, a rather slow convergence is inherent to
the usual iterative method of solution of the gap equation.
Since each iteration requires to
solve these integro-differential Bogolyubov equations,
a huge cpu time is necessary to obtain
a reliable solution in this way.

To make the procedure simpler, a new method
\cite{KKC} of solving the gap equation for
the case of nonlocal interaction is used (we refer to it as KKC).
This method was originally
devised for infinite nuclear matter but it is valid for any
situation provided
the gap $\Delta$ is much less than the Fermi energy
$\varepsilon_{\rm F}$.
The main computation problem of usual
iterative scheme originates from a
strongly nonlinear form of the integral gap equation with
a lot of iterations necessary to obtain the solution.
The main idea of the KKC method is that the nonlinear form
is important only for the magnitude of the gap
but not for its momentum dependence. As to the latter, it is determined
by integrals over a wide momentum space which
does not practically depend on
$\Delta$, provided the parameter $\Delta / \varepsilon_{\rm F}$ is small.
The gap $\Delta$ was identically represented in \cite{KKC} as a product
$\Delta(k) = \Delta_{\rm F} \chi(k)$ of the constant $\Delta_{\rm F}=
\Delta(k_{\rm F})$
and the "gap-shape" function $\chi(k)$ normalized to $\chi(k_{\rm F})=1$.
Then the initial nonlinear gap equation
was changed by a set of two equations. The first one,
for the gap-shape function $\chi(k)$, is a linear integral
equation which does not practically depend on the value of
$\Delta_{\rm F}$ and can be readily solved.
If the function $\chi(k)$ is known, the
equation for the gap amplitude  $\Delta_{\rm F}$
is just an algebraic nonlinear one which
can be solved straightforwardly with standard methods.

In the zero approximation of the KKC method,
the gap-shape function $\chi^{(0)}(k)$ is calculated with
$\Delta_{\rm F}=0$. Then the zero approximation for the gap magnitude
$\Delta^{(0)}_{\rm F}$ is readily found. At the next approximation,
the value of $\Delta^{(0)}_{\rm F}$ is used as an input for calculating
$\chi^{(1)}(k)$ and the procedure can be repeated. In principle,
the KKC method contains no additional approximations in comparison
with the standard one. It just rearranges the iterative scheme,
making the convergence much faster. Already the zero approximation
of the KKC method gives, as a rule, sufficiently accurate
results \cite{KKC,K1}.

The KKC method can be readily extended to the case of nonzero
temperature $T$ \cite{KKC,K1}.
In this case the gap-shape function is practically $T$-independent.
The reason is that at small temperature which is of interest
(the critical temperature is $T_c \simeq 0.5 \Delta_{\rm F}$) redistribution
of particles occurs in very narrow region nearby the Fermi surface
which does not practically influence $\chi({\bf k})$.
Especially simple result appears in the zero approximation
of the KKC method:
\beq %1
\Delta ({\bf k},T)=\Delta_{\rm F}(T) \,\chi({\bf k}).
\eeq

In the semi-infinite matter an additional coordinate dependence
appears, but all physical reasons of eq.~(1)-type
representation remain valid with the coordinate and momentum
dependent, but $T$-independent, shape function $\chi$ and
the  only $T$-dependent factor $\Delta_{\rm F}(T)$.
The latter one can be found as the asymptotical value inside
nuclear matter which coincides with that of infinite system
and is obtained by a much more simple calculation.
To find the shape function we go to the limit
$T \to T_c$ where  the gap equation becomes linear.
Such a method is used in this paper.
It turns out to be, in the case of semi-infinite nuclear matter,
much more convenient than the direct
solution of the gap equation.

In symbolic notation, the gap equation has the form
\cite{AB,Schuck}:
\beq %2
\Delta(T)=\Veff\,A_0^s (T)\,\Delta(T),
\eeq
where $\Veff$ is the effective pairing interaction
acting in the model space in which the superfluid two-particle
propagator $A_0^s $ is defined.

In the zero KKC approximation the temperature dependence of the gap
operator $\Delta({\bf k},T)$ in homogeneous nuclear matter
can be separated from its momentum dependence in the form of eq.~(1).
The shape function $\chi({\bf k})$ of the gap operator can be considered as
a $T$-independent one up to the critical temperature $T_c$.
If $T$ is close to $T_c$, the gap $\Delta$ is negligible,
and the propagator $A_0^s(T_c)$ in eq.~(2) coincides with  that of the
nonsuperfluid system which will be denoted as
$A_0(T)$. Hence at $T=T_c$ eq.~(2) is
reduced to a homogeneous linear integral equation for $\chi({\bf k})$.
Its eigenfunction corresponding to the eigenvalue $\lambda=1$ is
the shape-function we search for.

Let us consider now the gap operator for semi-infinite nuclear matter
which is nonuniform in the $x$-direction. We assume
that the temperature dependence can be separated from
the shape factor in the form similar to eq.~(1) :
\beq %3
\Delta (x_1,x_2,k^2_{\perp},T) = \Delta_{\rm F}(T)
\,\chi(x_1,x_2,k^2_{\perp}),
\eeq
where ${\bf k}_{\perp}$ denotes the momentum in the
${\bf s}$-direction which is
perpendicular to the $x$-axis. We use here the effective
interaction calculated
in \cite{BLSZ4} for the separable form
\cite{Par1,Par2} of the Paris potential
with form factors $g_i(k^2)$ (i=1,2,3) depending on
the relative momenta of interacting particles.
In notations of \cite{BLSZ4}, it has the form:

\beq %4
{\Veff}(x_1,x_2,x_3,x_4;k^2_{\perp},k^{\prime 2}_{\perp}) =
\sum\limits_{ij}\Lambda_{ij}(X,X')\,
g_i(k^2_{\perp},x) \,g_j(k^{\prime 2}_{\perp},x'),
\eeq
where $X{=}(x_1{+}x_2)/2$, $X'{=}(x_3{+}x_4)/2$, $x{=}x_1{-}x_2$,
$x'{=}x_3{-}x_4$,
and $g_i(k^2_{\perp},x)$ stands for the inverse Fourier transform
 of the form factor $g_i(k^2_{\perp}{+}k^2_x)$ in the $x$-direction.
It is obvious that the gap-shape factor can be also written as a sum
\beq %5
\chi(x_1,x_2;k^2_{\perp}) =
\sum\limits_i\chi_i(X)\,g_i(k^2_{\perp},x).
\eeq

After substitution of eqs.~(3)\,--\,(5) into eq.~(2) at
$T{=}T_c$ we arrive at
the following equation for the components $\chi_i$:
\beq %6
\chi_i(X)=\sum\limits_{lm}\int dX_1dX_2\,
\Lambda_{il}(X,X_1)\,B^0_{lm}(X_1,X_2,T_c)\,
\chi_m(X_2),
\eeq
where
\begin{eqnarray} %7
\lefteqn{
B^0_{lm}(X_1,X_2,T)=\int\frac{d{\bf k}_{\perp}}{(2\pi)^2}\,
dx_1\,dx_2\,g^*_l(k^2_{\perp},x_1)\,
g_m(k^2_{\perp},x_2)\times \qquad\qquad\qquad
}
\nonumber \\
& &{}\times A_0(X_1{+}{x_1\over 2},X_1{-}{x_1\over 2},
X_2{+}{x_2\over 2},X_2{-}{x_2\over 2};
k^2_{\perp},T).
\end{eqnarray}

As is known \cite{3auth}, for the two-particle propagator $A_0(T)$
at $T>0$ the Matsubara technique leads to the same expression as
at $T{=}0$:
\beq %8
A_0({\bf r}_1,{\bf r}_2,{\bf r}_3,{\bf r}_4,E=2\mu;T)=
\sum\limits_{\lambda, \lambda' } \,
 \frac {1{-}N_{\lambda}{-}N_{\lambda'}}
{2\mu {-} \varepsilon_{\lambda} {-}\varepsilon_{\lambda'}} \,
\varphi_{\lambda }({\bf r}_1)\varphi^*_{\lambda}({\bf r}_3)\,
\varphi_{\lambda'} ({\bf r}_2)\varphi^*_{\lambda'}({\bf r}_4),
\eeq
with $T$-dependent occupation numbers
\beq %10
N_{\lambda}(T)=\frac{1}{1 {+} e^{(\varepsilon_{\lambda}{-}\mu)/T}},
\eeq
where $\mu$ is the chemical potential of the system.
The summation in eq.~(8) is carried out over the single-particle states
$\lambda = \{n,{\bf k}_{\perp}\}$ with functions
$\varphi_{\lambda} ({\bf r})=y_n(x)\,e^{i{\bf k}_{\perp}{\bf s}}$
and energies $\varepsilon_{\lambda} = \varepsilon_n + {k^2_{\perp}}/{2m}$.

After simple transformations we finally obtain
\begin{eqnarray} %10
\lefteqn{
B^0_{lm}(X,X',T)=-\sum\limits_{n_1n_2}
\int\frac{d{\bf k}_{\perp}}{(2\pi)^2}\,\frac
{1{-}N_{n_1}(k_{\perp},T){-}N_{n_2}(k_{\perp},T)}
{\varepsilon_{n_1}{+}\varepsilon_{n_2}{+}k^2_{\perp}/m{-}2\mu}
\times
}\qquad\qquad\qquad \qquad\qquad\qquad
\nonumber \\
& &{}\times g^l_{n_1n_2}(k^2_{\perp},X)\,
g^m_{n_1n_2}(k^2_{\perp},X'),
\qquad\qquad
\end{eqnarray}
where
\beq %11
g^l_{n_1n_2}(k^2_{\perp},X)=\int g_l(k^2_{\perp},x)\,
y_{n_1}(X{+}{x\over 2})\,y_{n_2}(X{-}{x\over 2})\,dx.
\eeq

The effective interaction $\Veff$ \cite{BLSZ4} is defined
in such a way
that the model space involves only the single-particle states with
negative energies. Therefore the summation over $n_1,n_2$ and
integration over ${\bf k}_{\perp}$ in eq.~(10) is limited by the condition
$\varepsilon_{\lambda_1}, \varepsilon_{\lambda_2} <0$.

The above formulae determine the kernels of eq.~(6)
for the gap-shape function $\chi(x_1,x_2,k^2_{\perp})$. Solution of
this equation should be substituted into eq.~(3) together with
the factor of $\Delta_{\rm F}(T){=}0$ which can be found at
$X_{12}{\to}-\infty$.
Therefore it coincides with that of infinite nuclear matter.

Before going over to solution of the above equations for the case
of semi-infinite nuclear matter, let us check the accuracy
of eq.~(1), i.e. the zero approximation of the KKC method,
in infinite matter for the specific kind of the $NN$-potential in
the separable form \cite{Par1,Par2}.
It was used for studying pairing
in neutron and nuclear matter within the Brueckner theory
in Refs. \cite{BL1,BL2}. We present here some results of these
calculations in the form which is convenient for analysis of
the $T$-dependence of the gap-form function.
For the separable potential under consideration the momentum
dependence of $\Delta$ is given by coefficients $C_i$ of the sum
\beq %12
\Delta(k,T) =
\sum\limits_i C_i(T) \, g_i(k^2).
\eeq

In fact, it is sufficient to check the $T$-independence of the
ratios of these coefficients to each other.
It results that
the $T$-independence of the ratios of $C_2/C_1$ and $C_3/C_1$
is really true for both neutron and nuclear matter
for different densities up to $T=T_c$ within 2\%$\div$3\%- accuracy.
An example for nuclear matter at $k_{\rm F}=0.8$ fm$^{-1}$
which corresponds to the maximum value of $\Delta$
is given in Table 1.

Let us return to consideration of semi-infinite matter.
Eq.~(6) can be considered as a particular case of a more
general set of homogeneous integral equations
\beq %13
\chi_i(X) = \lambda(T) \, \sum\limits_{j} \int dX' \,
 K_{ij}(X,X';T) \, \chi_j(X),
\eeq
where $K_{ij}(X,X';T)$ are the kernels of this equation
for arbitrary $T$ and $\lambda(T)$ stands for the eigenvalue.
The critical temperature can be found from
the condition that the minimum
eigenvalue $\lambda_1(T_c)$ is equal to unit. In principle, such
a way which  involves solution of eq.~(13) for a number of values of $T$
is possible but rather cumbersome. In the situation under consideration,
when the pairing in infinite nuclear matter exists, it is obvious that $T_c$
in semi-infinite matter is the same as in infinite one. Therefore
it is much simpler to find  $T_c$ for infinite matter and use it in eq.~(6)
(or eq.~(13) at $\lambda{=}1,\; T{=}T_c$) as an input.

\newpage
\centerline{\bf Table 1}
\vskip .15 cm

\centerline{Demonstration of the $T$-independence of the gap-shape}
\centerline{function for nuclear matter at $k_{\rm F}=0.8$ fm$^{-1}$.}
\vskip 10pt
\begin{center}
\begin{tabular}{|l|c|c|}
\hline
$T$, MeV \rule[-9pt]{0pt}{27pt}& $ C_2/C_1$& $ C_3/C_1$  \\
\hline
0  \rule{0pt}{16pt} & 0.355 & $-$0.0271  \\
0.25                & 0.355 & $-$0.0271  \\
0.50                & 0.355 & $-$0.0271  \\
0.75                & 0.354 & $-$0.0271  \\
1.00                & 0.353 & $-$0.0270  \\
1.25                & 0.350 & $-$0.0270  \\
1.30                & 0.349 & $-$0.0269  \\
1.35                & 0.349 & $-$0.0269  \\
1.40                & 0.348 & $-$0.0269  \\
1.45                & 0.347 & $-$0.0269  \\
1.50                & 0.347 & $-$0.0270  \\
1.55($\simeq T_c$)  & 0.348 & $-$0.0271  \\
\hline
\end{tabular}
\end{center}
\vskip .15cm

Let us now describe briefly how to calculate
the set of integral equations, eq.(6). The kernels $K_{im}(X,X')$ of these
equations are given by folding of the coefficients
$\Lambda_{il}(X,X_1)$ of the effective
interaction $\Veff$ with those $B^0_{lm}(X_1,X';T_c)$
of the propagator $A_0(T{=}T_c)$.
The first ones were calculated in \cite{BLSZ4} for the value of chemical
potential $\mu=-8$ MeV which simulates the situation in finite nuclei.
In this case the critical temperature is $T_c{=}0.124$ MeV.
In fact, the interval $\{X\}=(-8\,{\fermi}, 8\,{\fermi})$
was considered only because
properties of $\Veff$ are trivial outside this interval.
So, at $X,X'<-4$ fm the magnitude of $\Veff$
coincides practically with that calculated for infinite matter,
whereas at $X,X'>4$ fm
it tends rapidly to the free $T$-matrix taken at negative
two-particle energy $E=2\mu$.
As it is shown in \cite{BLSZ4}, at a fixed value of $X_0{=}(X{+}X')/2$
the effective interaction $\Veff(X,X')$ vanishes rapidly when the
relative distance $t{=}X{-}X'$ is growing, so that integrals involving
$\Veff(X,X')$ can be cut at $|t|>4$ fm. That is why the interval
$(-8\,\fermi, 8\,\fermi)$ for the effective interaction
is sufficient for any calculations.
As to eq.(6) for the gap-form function,
a much wider $X$-space should be considered,
$\{X\}=(-L_{\inn},L_{\exx})$ with a minimum value
of $L_{\inn} \simeq 40$ fm
(the value of $L_{\exx} = 8$ fm is large enough). The reason
is that at small value of
$\Delta_{\rm F}\lsim 1$ MeV the correlation length
$\xi\sim v_{\rm F}/\Delta_{\rm F}$ is very big ($\xi > 10$ fm).
The distance between the left cut-off $L_{\inn}$ and the point under
consideration (say, $X\gsim -10$ fm) should exceed
a correlation length for the effects of the left boundary not to
distort the asymptotic behaviour of the gap-shape function
inside nuclear matter essentially.
For $X,X'<-8$ fm we use the effective interaction
${\cal V}_{\mbox{\scriptsize eff}}^{\infty}(t)$
calculated for infinite nuclear matter. It is obvious that it depends only
on the difference $t{=}X{-}X'$.

The propagators $B^0_{lm}(T)$ are new ingredients of the problem.
They can be readily calculated in accordance with formulae (10),~(11)
using the technique elaborated in \cite{BLSZ4}.

Such a large $X$-space makes it difficult to solve eq.~(6)
in the coordinate space directly along the way which was
used in Ref.~\cite{BLSZ4} for a similar equation for $\Veff$.
It is more convenient to use the Fourier expansion of the gap-shape
function in the interval $(-L_{\inn},L_{\exx})$:
\beq %14
\chi_i(X) = \sum\limits_n \, \chi_i^n f_n(X)
\eeq
where $f_n(x)$ are $\sin(2\pi n(X{-}X_c)/L)$ and
$\cos(2\pi n(X{-}X_c)/L)$, $L{=}L_{in}{+}L_{ex}$,
$X_c=(L_{ex}-L_{in})/2)$.
The kernels $K_{ij}(X,X')$ are also expanded in the double
Fourier series. Finally, we deal with a set of
homogeneous linear equations for the coefficients $\chi_i^n$ :
\beq %15
\chi_i^n= \sum\limits_{j=1}^3 \sum\limits_{n'=1}^N
\, K_{ij}^{nn'}\, \chi^{n'}_j,
\eeq
with the matrix $K_{ij}^{nn'}$ having not very high dimension.
In this calculation we used the interval of
$\{X\}=(-40\,\fermi,10\,\fermi)$ ($X_c{=}-15$ fm),
the value of
$N{=}101$ being enough to obtain the
accuracy better than 1\%. When the coefficients  $\chi_i^n$ are found,
the gap-shape function $\chi(X,t,k^2_{\perp})$
can be readily obtained with use of eqs.
(14) and (5). To obtain the complete gap at zero temperature
$\Delta (X,t,k^2_{\perp},T=0)$, one has  just to multiply  $\chi$
by a constant $\Delta_{\rm F}=0.167 $ MeV found for infinite nuclear
matter.

To present the results in a more transparent way,
it is convenient to calculate the zero moment of $\chi(X,x,{\bf s})$
over the relative coordinates:
\beq %16
\chi_0(X) = \int d{\bf s} \int dt \,
\chi(X+t/2,X-t/2,{\bf s} )
= \sum\limits_i \, \chi_i(X).
\eeq
Here, in accordance with Ref. \cite{BLSZ4}, the normalization
$g(k^2=0)=1$ was used.
Let us also consider the function
\beq %17
\chi_{\rm F}(X) =  \sum\limits_i \, \chi_i(X) \,
g_i(k^2=k_{\rm F}^2(X) ),
\eeq
where $k_{\rm F}(X)=\sqrt{2m(\mu{-}U(X))} $ is the local Fermi
momentum ($k_{\rm F}(X)=0$ for\\ $\mu{-}U(X) <0 $).
The corresponding moments of the gap are as follows:
$\Delta_{0}(X)=\Delta_{\rm F} \chi_0(X)$,
$\Delta_{\rm F}(X)=\Delta_{\rm F} \chi_{\rm F}(X)$.
The latter function
is very important for the interpretation of results. It
gives approximately matrix elements of $\Delta$ for the single-particle
states taken at the Fermi surface. Analysis of this approximation
for the case of the semi-infinite geometry is given in Ref. \cite{BLSZ4}.
It should be noted that this type of the local approximation is
commonly used for description of pairing in finite nuclei \cite{Sch2}.
Its accuracy for neutron stars was examined in \cite{Brgl}.
It worth to stress that in our paper it is used
only for a more transparent presentation of results
which were obtained without this approximation.

Similar calculations were repeated for the chemical potential
$\mu=-16$ MeV which is the "self-consistent" value for semi-infinite
nuclear matter coinciding with that of infinite matter.
In this case we have $T_c=0.57$ MeV and $\Delta_{\rm F}=0.94$ MeV.

Both functions, $\Delta_{0}(X)$ and $\Delta_{\rm F}(X)$,
are drawn in Fig.1 for both values of $\mu$.
As it is seen, all the curves possess pronounced surface maxima.
Such a surface behaviour
appears due to a big surface value of the effective
pairing interaction \cite{BLSZ4}.

For more convenient analysis of the $\mu$-dependence of the surface
effect, the gap-shape functions  $\chi_{\rm F}(X)$
for both values of $\mu$ are drawn together in Fig.2.
As it is seen,  the surface effect in $\Delta$
is more pronounced for $\mu=-8$ MeV.
Indeed, in this case the ratio of the maximum
surface value to the asymptotic one inside the matter is $\simeq$1.8,
whereas it is $\simeq$1.5 for $\mu=-16$ MeV.
Besides that, the position of the maximum is
closer to the surface ($X=0$) for $\mu=-8$ MeV than for $\mu=-16$ MeV.

It should be noted that the surface effect for $\Delta$
turns out to be less pronounced than the one for the effective
interaction itself. This is a consequence of a strong
surface-volume coupling which takes place in the gap-equation
due to a big pairing correlation length which is significantly
larger than the width of the surface layer. Such a coupling is
inherent to a pure quantum calculation and can be partially
lost when a local approximation is used. It suppresses partially
the surface maximum in $\Delta$ and makes the $\mu$-dependence
of the surface effect smoother.

Such a dependence of the gap on $\mu$
originates from two reasons. The first one is a direct energy dependence
of the effective interaction $\Veff$ taken at $E=2\mu$ \cite{BLSZ4}.
The second one is the strong momentum dependence of the form-factors
$g_i(k^2)$ taken at $k^2=k^2_{\rm F}(X)={2m(\mu{-}U(X))}$. Both effects
cooperate in enhancing the surface values of the
effective interaction and of $\Delta$ for smaller values of $\mu$.
However, there is an effect which works in the opposite direction
making the $\mu$-dependence of the surface maximum
of $\Delta$ less pronounced.
This is a decrease of $\Delta_{\rm F}$ at smaller values of
$|\mu|$ which results in growing the correlation length. The latter
smooths the surface effect itself and its $\mu$-dependence.

The energy dependence of pairing should be
important for consideration of the nuclear drip-line where the chemical
potential $\mu $ tends to zero. The analysis of situation at small
values of $\mu$ is in progress.

%This research was partially supported
%by Grant No.~96-02-06254 from the Russian
%Foundation for the Fundamental Research.
The authors thank V.~A.~Khodel for the advice to use
the idea of temperature independence of the gap-form function
to calculations for semi-infinite nuclear matter.
They wish also to acknowledge S.~T.~Belyaev, S.~A.~Fayans
and  P.~Schuck for fruitful discussions.
Two of us (E.~S. and M.~Z.) thank
INFN and Catania University for hospitality during their stay
in Catania.

\newpage

\centerline {\bf Figure captions.}

\vskip 0.7 cm

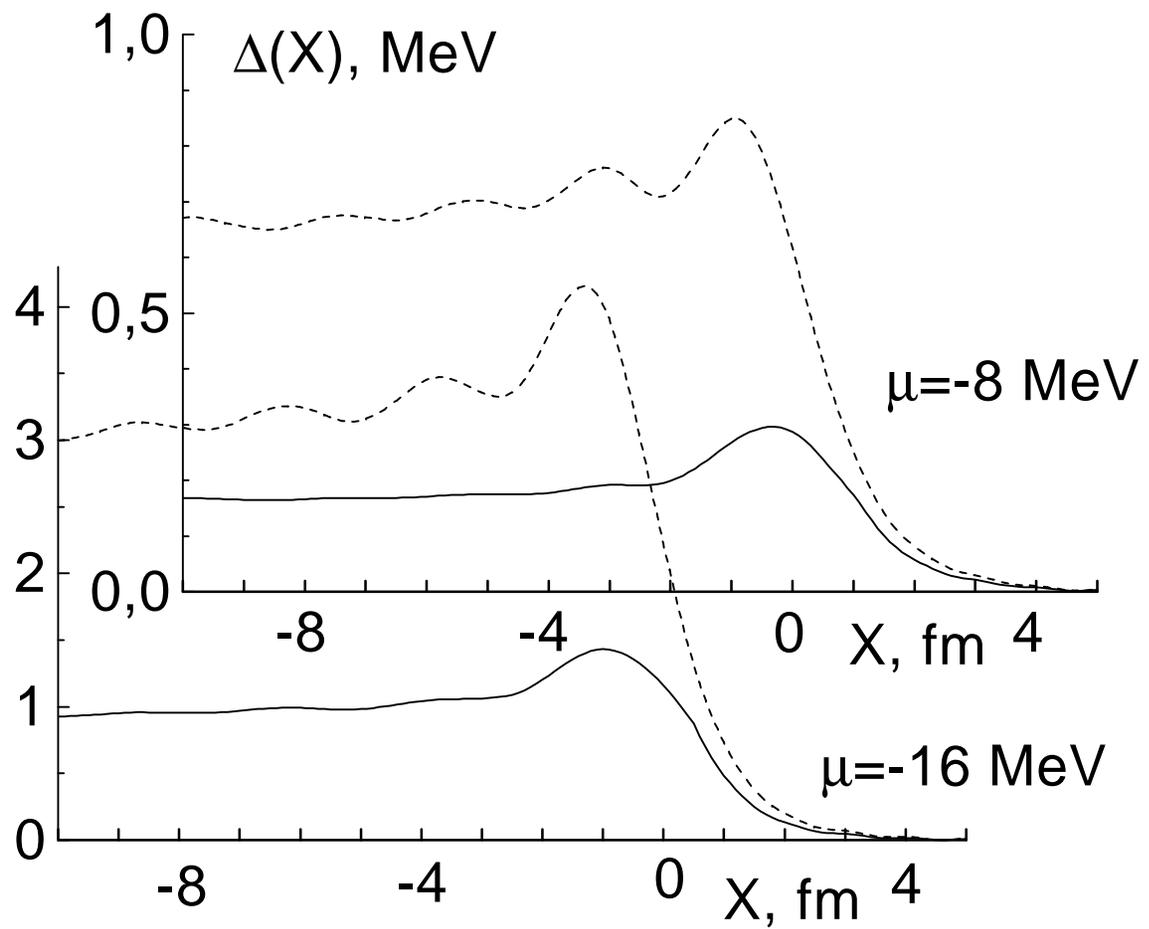
Fig.~1. The zero moment $\Delta_0(X)$ of the gap (dashed lines),
and the function $\Delta_{\rm F}(X)$ (solid lines) for two values
of $\mu$.

\vskip 0.2 cm
Fig.~2. The gap-shape $\chi_{\rm F}(X)$ for $\mu{=}-8$MeV (dashed lines),
and for $\mu{=}-16$MeV (solid lines).

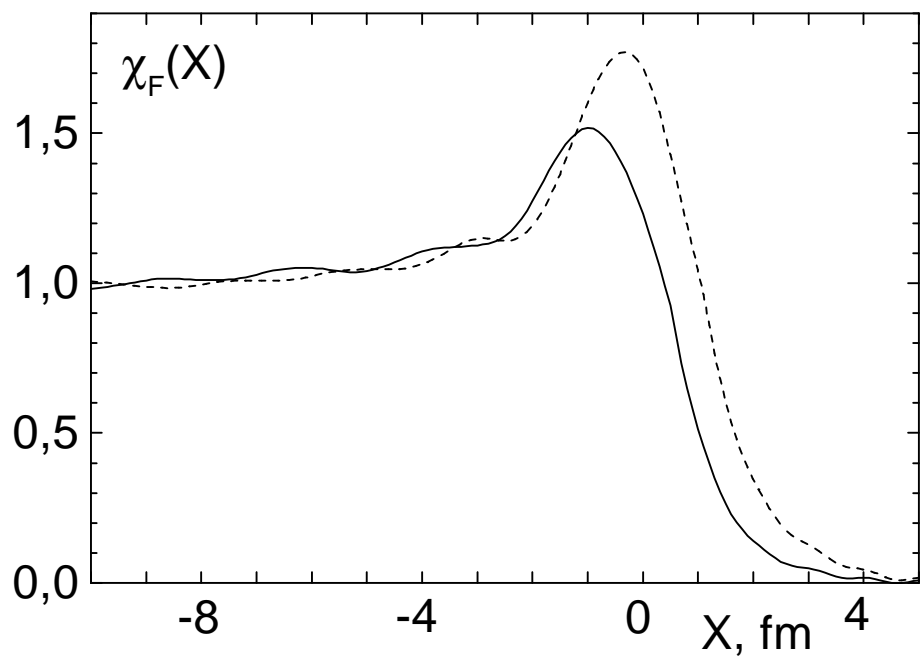
\end{document}